\documentclass[aps,prf,twocolumn,floats,amsmath,amssymb,longbibliography,showpacs,floatfix]{revtex4-1}
\usepackage[dvipsnames,rgb,dvips]{xcolor}
\usepackage{graphicx,overpic}
\usepackage{siunitx}
\usepackage{bbold}
\usepackage{mathrsfs}

\newcommand{\ve}[1]{\ensuremath{\mbox{\boldmath$#1$}}}
\newcommand{\ma}[1]{\ensuremath{\mathbb{#1}}}

\newcommand{\st}{\text{St}}

\begin{document}
\title{Bifurcations in droplet collisions}
\author{A. Dubey$^1$, K. Gustavsson$^1$, G. Bewley$^2$, and B. Mehlig$^{1}$}
\affiliation{
\mbox{}$^1$Department of Physics, Gothenburg University, 41296 Gothenburg, Sweden\\
\mbox{}$^2$Sibley School of Mechanical and Aerospace Engineering,
Cornell University, USA}

\begin{abstract}
Saffman and Turner (1957) argued that the collision rate for droplets in turbulence increases as the turbulent strain rate increases.  But the numerical simulations of Dhanasekaran {\em et al.} (2021) in a steady straining flow show that the Saffman-Turner model is oversimplified because it neglects droplet-droplet interactions. These result in a  complex dependence of the collision rate on the strain rate and on the differential settling speed.  Here we show that this dependence is explained by a sequence of bifurcations in the collision dynamics.  We compute the bifurcation diagram  when strain is aligned with gravity, and show that it yields important insights into the collision dynamics.  First, the steady-state collision rate remains non-zero in the limit ${\rm Kn}\to 0$, contrary to the common assumption that the collision rate tends to zero in this limit (Kn is a non-dimensional measure of the mean free path of air). Second, the non-monotonic dependence of the collision rate on the differential settling speed is explained by a grazing bifurcation.  Third, the bifurcation analysis explains why so-called \lq closed trajectories\rq{} appear and disappear. Fourth, our analysis predicts  strong spatial clustering near certain saddle points, where the effects of strain and differential settling cancel.  
\end{abstract}
\maketitle

\section{Introduction}

Turbulent aerosols are suspensions of particles or droplets in a turbulent flow. Common examples are droplets  or ice crystals in atmospheric clouds \cite{Pru10}, and dust particles in circumstellar accretion disks \cite{Anders}. The physical properties of such systems are determined by 
collisions between the particles \cite{wilkinson2008stokes,Gus14c}.  In clouds, turbulent strains bring droplets into close contact \cite{saffman1956collision}. Apart from this mechanism,
their collision dynamics is 
influenced by gravity, hydrodynamic forces, and direct electrostatic interactions if the droplets are charged~\cite{magnusson2021collisions}. Fluid-mediated, hydrodynamic  interactions tend to  bend the paths of two approaching droplets around each other \cite{Kim2005}, potentially preventing collision. But when the droplets are  very close -- when their interfacial separation is smaller than the mean free path of air --  the hydrodynamic  approximation breaks down, reducing the hydrodynamic repulsion of approaching droplets  \cite{Sun96}. Another factor that affects collision rates is particle  inertia, which allows the droplets to detach from the flow, increasing the collision rate \cite{Gus16}. Larger droplets
may accelerate the surrounding fluid significantly, so that convective  fluid inertia must be taken into account for similar-sized droplets  \cite{Kle73}.

In other words, the collision dynamics of droplets in turbulence is quite complex, and though it has been studied for more than 50 years, it is not known  how 
to parameterise the collision rate in terms of  the non-dimensional parameters: the Stokes number St (particle inertia), 
the non-dimensional differential settling speed $Q$ (gravitational settling), turbulence intensity (strength of the turbulent strains), Reynolds and Strouhal numbers (fluid inertia), the Coulomb number (electrostatic interactions), the Knudsen number Kn (breakdown of the hydrodynamic  approximation at distances smaller than the mean free path of air), 
and the ratio of droplet radii. 

\citet{saffman1956collision} computed the effect of turbulent strain on the collision rate, assuming that the droplets follow the flow, neglecting all other interactions between them.  They found that
the collision rate for droplets of radii $a_1$ and $a_2$ increases with strain $s$ as 
$\mathscr{R}_s\sim n_0 (a_1+a_2)^3 s$, where $n_0$ is the particle-number density. They compared with the collision rate of small spherical droplets settling under gravity in a quiescent fluid, $\mathscr{R}_g\sim n_0 v_s (a_1+a_2)^2$ where 
$v_s = \tfrac{2\rho_p}{9\rho_f}(a_{2}^2-a_{1}^2)g/\nu$ is the differential settling speed with viscosity $\nu$, gravitational acceleration $g$, and ratio of  droplet and air mass densities 
$\rho_p/\rho_f \gg 1$. 
Their main conclusions are that turbulence facilitates collisions between droplets of similar sizes, 
 and that the collision rate increases as the turbulent strain $s$ increases. However,  Fig.~10 of Ref.~\cite{Dhanasekaran2021} indicates that this picture may be oversimplified. 
\citet{Dhanasekaran2021}  computed the collision dynamics of small spheres settling in a steady straining flow, taking into account hydrodynamic interactions and the breakdown of the hydrodynamic  approximation at small separations. Effects of particle and fluid inertia, electrostatic interactions, and unsteady turbulent fluctuations were not considered. 
Their numerical results show that the collision rate depends sensitively on the non-dimensional differential settling speed $Q$, a measure of the relative strength of strain and gravity.

Here we show that the observed dependence of the collision rate  upon $Q$  reflects qualitative changes -- {\em bifurcations} -- in the collision dynamics as  $Q$ is varied.
We analysed the bifurcations for a specific example~\cite{Dhanasekaran2021}: two droplets of different sizes settling in a steady straining flow. We chose a straining flow aligned with the direction of gravity (Fig.~\ref{fig:schematic}). 
We found the equilibria  of the collision dynamics, and we determined how they appear, disappear, and how their stabilities and  invariant manifolds change as the parameter $Q$ is varied.

The bifurcation analysis allows the following conclusions. 
First, for a certain range of $Q$ values, the steady-state collision rate becomes independent of Kn for small Knudsen numbers, contrary to the common 
assumption that droplets cannot collide in this limit \cite{Hocking1973,Davis1984,Pru10, Pum16}. 
Second, a global grazing bifurcation (called \textcircled{\footnotesize $\alpha$} below)  causes a sensitive dependence of the collision rate upon $Q$.
Third, our bifurcation analysis allows to understand the significance of so-called \lq closed trajectories\rq{} that appear and disappear as $Q$ is varied.
These are trajectories that start and end on the collision sphere \citep{Batchelor1972,Zeichner1977,Brunk1998,Dhanasekaran2021}.
Fourth, we find strong spatial clustering near certain saddle points, on their unstable invariant manifolds. In our model this leads to a divergent pair correlation function
at separations of the order of several droplet radii.

Our main conclusions rely on the fact that  the collision dynamics has a boundary at $R=a_1+a_2$. The global grazing bifurcation occurs
when an invariant manifold collides with this boundary~\cite{diBernardo2008,Brogliato2016}.
Grazing bifurcations are  usually discussed for piecewise smooth dynamical systems \cite{Foale1994,dibernado2010discontinuity,Brogliato2016}. In such systems boundaries occur between the  smooth parts.
 In our case, however, the radial derivative of the relative droplet velocity diverges  at the boundary  $R \to a_1+a_2$, so our system is not piecewise smooth.
This is the origin of the sensitive dependence of the collision rate upon Q mentioned above.

 \section{Model}
\begin{figure}[t]
\begin{overpic}[scale=0.4]{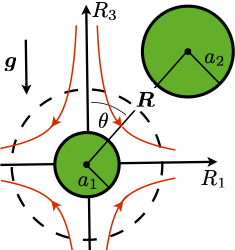}
\end{overpic}
\caption{\label{fig:schematic}  Two droplets settling in a straining flow (thin red lines), shown in the rest  frame of the smaller droplet, with radius $a_1< a_2$. 
The compressive direction of the strain  is  aligned with gravity $\ve g$.  Since the system has rotation symmetry about the $R_3$-axis, 
 it suffices to analyse the dynamics in the $R_1$--$R_3$ plane.
}
\end{figure}

We begin with the definition of the non-dimensional parameters St and $Q$. The Stokes number is defined as St$\;=s \tau_p$ with strain rate $s$ and
particle response time $\tau_p = (2\rho_p/9\rho_f) (\overline{a}^2/\nu)$, where $\overline{a}= (a_1+a_2)/2$ is the mean particle radius. The effect of gravity
is parameterised by the non-dimensional differential settling speed $Q=v_s/(\overline{a}s)$.
Here we consider droplets small enough so that we can take the limit $\st\to 0$. 
At the same time we assume that  $Q$ remains finite. 
In this overdamped limit, the relative dynamics is just given by $\dot{\ve R} = \ve V(\ve R)$
where $\ve R=\ve x^{(2)}-\ve x^{(1)}$ is the spatial separation between two particle centres, and
$\ve V = \ve v^{(2)}-\ve v^{(1)}$ is their relative velocity. This limit neglects 
singularities and multi-valued particle velocities in the inertial  collision dynamics due to caustics \cite{Fal02,Wil05,Wil06}. The probability of observing two droplets 
at separation $\ve R$ is given by the continuity equation 
\begin{align}
\label{eq:ce}
 \frac{\partial P(\ve R,t) }{\partial t} + \nabla \cdot [\ve V(\ve R) P(\ve R,t)]=0\,,
\end{align}
with the boundary condition $P(\ve R,t) \to 1$ as $R \to \infty$.  This boundary condition reflects the uniform spatial distribution at large separations of  small droplets in an incompressible flow. 
The collision rate is given by \cite{Gus08}
\begin{align}
\label{eq:collision_rate}
\mathscr{R}_t&\!=\!-\!\!\!\!\lim_{R\to a_1\!+\!a_2} \!\!n_0\!\int\!{\rm d}\Omega \,V_R(\ve R)  P(\ve R,t) \Theta\big[-V_R(\ve R)\big]\,,
\end{align}
where $V_R = \ve V\cdot \hat{\ve R}$ is the radial relative velocity, $R=|\ve R|$ is the centre-of-mass distance between the droplets,
$\Omega$ is the solid angle, and $\Theta$ is the Heaviside function.

Following Ref.~\cite{Dhanasekaran2021} we consider a steady straining flow  with velocity $\ve u(\ve x)= 
\tfrac{1}{2}(sx _1\hat{\bf e}_1+sx _2\hat{\bf e}_2-2s x_3 \hat{\bf e}_3)$, where $\hat{\bf e}_i$ denotes the unit vector in direction $R_i$ (Fig.~\ref{fig:schematic}). The strain-rate matrix $\ma S$ is diagonal with entries $S_{11}=s/2, S_{22}=s/2, S_{33}=-s$. 
We assume that the compressive axis of the strain is aligned with the gravitational acceleration (Fig.~\ref{fig:schematic}). 

In the overdamped limit, the relative velocities
 $\ve V(\ve R)$  are computed using the 
 hydrodynamic mobility tensor that relates particle velocities to the given external forces  in a linear flow \cite{Kim2005}.  The elements of this tensor depend on the particle radii $a_1$ and $a_2$, and on the particle separation $\ve R$. 
In non-dimensional variables ($\ve R \to\ve R/\overline{a}$, $t\to ts$) the equations of motion read \cite{Dhanasekaran2021}
\begin{align} \label{eq:eom}
\dot{R}_i &= V_i \nonumber\\ 
 V_i &= S_{ij} R_j -\left[A \frac{R_i R_k}{R^2} + B\left(\delta_{ik}-\frac{R_i R_k}{R^2}\right)\right] S_{kl}\,R_l \nonumber\\
 &-\left[ L\frac{R_i R_k}{R^2} + M\left(\delta_{ik}-\frac{R_i R_k}{R^2}\right)\right] \, Q \, \delta_{k3}\,,
\end{align}
where summation over repeated indices is implied. Here $A$, $B$, $L$, and $M$ are mobility functions, formed using elements of the mobility tensor mentioned above. These mobility functions depend only upon the droplet separations and their radius ratio. The mobility functions are not known in closed form. 
To evaluate $A$, $B$, $L$, and $M$, we use series expansions in $a/R$ derived in Refs.~\cite{Jeffrey1984,Townsend2018,Wang1994}.
In this hydrodynamic approximation, the radial mobilities  $1-A$ and $L$ decay linearly to zero as the interfacial separation between the droplets vanishes. 
However, when the interfacial separation between the droplets becomes comparable to the mean free path of air, $\ell$, 
the hydrodynamic approximation breaks down. This gives rise to non-hydrodynamic effects, parameterised by the Knudsen number ${\rm Kn}=\ell/\overline a$. \citet{Sun96} derived how these  effects change the functional form of the mobility functions on close approach. \citet{Dhanasekaran2021} showed how to match the radial non-continuum mobilities $A$ and $L$ 
 to the the series expansions of their hydrodynamic counterparts. We followed this approach, and like Ref.~\cite{Dhanasekaran2021} we did not 
 account for Kn-corrections to  the tangential mobilities $B$ and $M$. While non-continuum effects may change how the mobilities approach their values upon contact \cite{li2021non}, it is expected that  this change has  a small effect on the collision dynamics~\cite{Dhanasekaran2021}, at least when the droplet radii are not too different.

There are two reasons why the collision dynamics \eqref{eq:eom} is not smooth. First, the mobility functions are non-smooth at $R=2$. The radial mobilities decay to zero as $(\log \log \tfrac{\text{Kn}}{R-2})^{-1}$ as $R\to2$, and the radial derivative of this expression diverges in this limit. This means that the dynamics is not unique once it reaches $R=2$. 
The tangential mobilities behave as $(\log \tfrac{1}{R-2})^{-1}$ near $R=2$. The non-smoothness of the mobilities means that bifurcations at $R=2$ need not have the normal forms known for smooth dynamical systems \cite{Strogatz2000,Ott02}.  Second, we assume that the droplets coalesce upon collision,
so that the dynamics arrests on the collision sphere, at $R=2$. The collision sphere therefore acts as a boundary, allowing global grazing bifurcations \cite{diBernardo2008,Brogliato2016} to occur.

\section{Results}

\begin{figure*}[p]
\begin{overpic}[scale=0.275]{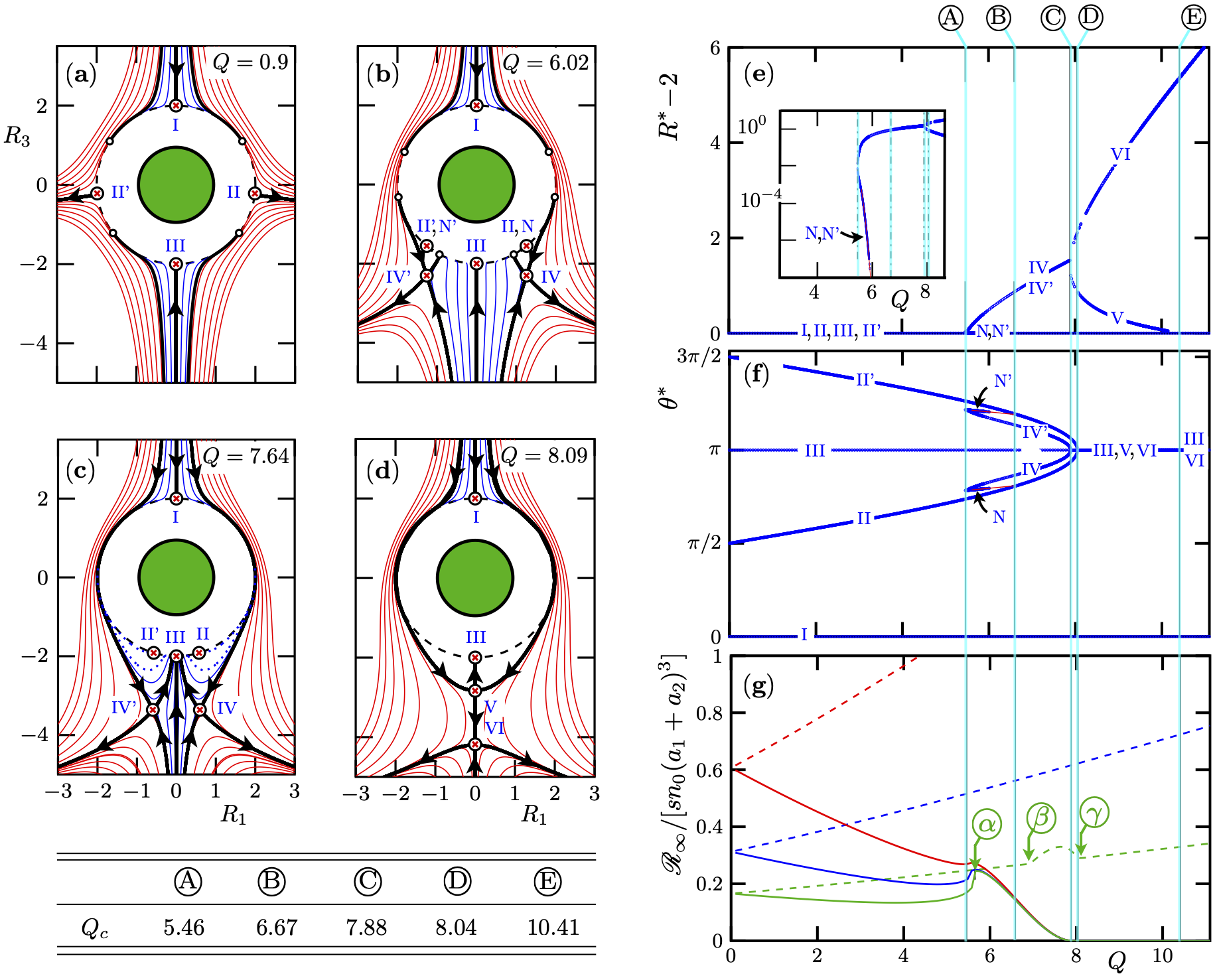}
\end{overpic}
 \caption{\label{fig:phase1} ({\bf a}-{\bf d}) Phase portraits of the collision dynamics for different values of the non-dimensional differential settling speed $Q$. Trajectories leading to collision (blue), no-collision (red),  invariant manifolds and grazing trajectories (solid black), collision sphere (dashed). 
 In panel ({\bf c}) closed trajectories that begin and end on the collision sphere are plotted as dotted blue lines. There are such trajectories in panel ({\bf b})  as well, but they are not shown.
 Encircled red crosses denote fixed points. Small white circles on the collision sphere denote end points of grazing trajectories or invariant manifolds.
 ({\bf a})  $Q=0.9$, $a_1/a_2 = 0.9$, Kn$=10^{-3}$. ({\bf b})   $Q=6.02$.  ({\bf c}) $Q=7.64$.  ({\bf d})  $Q=8.09$.
In panels ({\bf c}) and ({\bf d}), the stable manifolds of IV and IV$'$ and V coming from $R_3=\infty$ are close to but do not touch the collision sphere. 
  ({\bf e},{\bf f})  Bifurcation diagrams for $a_1/a_2=0.9$ and Kn$=10^{-3}$, showing the locations $[R^\ast-2,\theta^\ast$] of the fixed points  I \textendash VI, N as well as their symmetric counterparts (primed).  Blue curves correspond to numerical data,  red  lines in   panel  ({\bf f})  to  Eq.~(\ref{eq:B9}).  The inset in panel  ({\bf e}) shows
  the same data, but with a logarithmic $y$-axis. The table
  lists the locations of
  the bifurcations  \textcircled{{\footnotesize A}}\textendash \textcircled{\footnotesize E}, also shown as vertical lines in the panels  on the right.  (\textbf{g}) 
  Steady-state collision rate $\mathscr{R}_\infty$  as a function of $Q$, for
  $a_1/a_2 = 0.9$, and for  Kn=$10^{-1}$ (red), $10^{-2}$ (blue), and $10^{-3}$ (green). Dashed lines show the collision rate due to droplets approaching from $R_3=+\infty$, while solid lines correspond to droplets approaching from $R_3=-\infty$.  Also shown are the locations of the three grazing bifurcations  \textcircled{{\footnotesize$\alpha$}}\textendash \textcircled{{\footnotesize$\gamma$}} 
  for Kn=$10^{-3}$. 
  } 
\end{figure*}

\subsection{Phase portraits}
Figure~\ref{fig:phase1} shows phase portraits of the collision dynamics. We plotted the droplet paths in the $R_1$-$R_3$ plane, in the rest frame of the smaller droplet. 
It suffices to consider the dynamics in the $R_1$-$R_3$ plane because the system has rotational symmetry about the 
$R_3$-axis when the compressive axis of strain is aligned with gravity. Panel ({\bf a}) shows the phase portrait 
for Kn$\;=10^{-3}$ and a small value of the non-dimensional settling velocity ($Q=0.9$) where strain dominates, and gravitational settling is of secondary importance. 
There are four fixed points, labeled I, II, III, and II$^\prime$.
  The phase portrait is symmetric under  the reflection $R_1\to -R_1$, and we label the symmetric counterpart of a fixed point
by a prime. The phase portrait shown in panel ({\bf a}) is very similar to that at $Q=0$, just slightly distorted. At $Q=0$, the relative-velocity field $\ve V(\ve R)$ is symmetric under reflections $R_3 \to - R_3$ and $R_1 \to -R_1$. Therefore $\ve V$ must be parallel to $\hat{\bf e}_3$ for $R_1=0$, and $\ve V\propto \hat{\bf e}_1$ for $R_3=0$.
Using that the radial velocity vanishes at $R=2$, we obtain
\begin{align} \label{eq:fp}
 \begin{bmatrix}2\\0
 \end{bmatrix}\, (\text{I}),\quad
 \begin{bmatrix}2\\\tfrac{\pi}{2}
 \end{bmatrix}\,(\text{II}),\quad
  \begin{bmatrix}2\\ \pi 
 \end{bmatrix}\,(\text{III}),\quad
  \begin{bmatrix}2\\ \tfrac{3\pi}{2}
 \end{bmatrix}\, (\text{II$^\prime$})
 \end{align}
for the locations $[R^\ast,\theta^\ast]$ of the four fixed points at $Q=0$.
Here $R\equiv|\ve R|$ and $\theta$ is the polar angle measured from the positive $R_3$ axis (Fig.~\ref{fig:schematic}).
 For non-zero values of $Q$, symmetry implies that $\theta^\ast$ remains at $0$ for I, and at $\pi$ for III. The precise locations of II and II$'$ on the collision 
 sphere are given in Appendix~\ref{sec:appendix_a}. The velocity field cannot be linearised on the collision sphere. Nevertheless, the flow near these fixed points resembles that near a saddle, with stable and unstable manifolds. 
Saddles I and III have stable manifolds on the $R_3$-axis, while II and II$^\prime$ have unstable manifolds on the $R_1$-axis. In addition, the fixed points are connected by manifolds on the collision sphere (not shown). 
Panel ({\bf a}) indicates that droplets approach along the stable  manifolds of saddle I and III, and escape along the unstable manifolds of II and II$^\prime$. 
The trajectory separating colliding trajectories from non-colliding ones
is a grazing trajectory: it is tangential to the collision sphere at the point where it touches the collision sphere (marked by a circle).

Panel ({\textbf b})  corresponds to $Q=6.02$ and ${\rm Kn}=10^{-3}$. This phase portrait is qualitatively different from that in panel ({\textbf a}). 
There are additional fixed points, suggesting that  bifurcations occurred. The  new fixed points are a stable node, N,  and a saddle point, IV, as well as their symmetric counterparts. 
Note that the node N is very close to the saddle point II, they cannot be distinguished in the figure. 
We see  that a grazing trajectory determines collision outcomes  for  paths approaching from $R_3 = +\infty$. But for paths approaching from $R_3 =-\infty$,
the separatrix between colliding and non-colliding trajectories is formed by stable manifolds of the saddle points~IV~and~IV$'$. 
One stable and one unstable manifold of the saddle IV intersect the collision sphere. The same is true for IV$'$. Between these stable and unstable invariant manifolds, there are closed trajectories 
 that start and end on the collision sphere \citep{Batchelor1972,Zeichner1977,Brunk1998,Dhanasekaran2021} (not shown).

Panel ({\bf c}) shows the collision dynamics for $Q=7.64$ and ${\rm Kn}=10^{-3}$. The nodes disappeared, and the stable manifold of the saddle IV which previously intersected the collision sphere now extends to $R_3 \to +\infty$. It does not touch the collision sphere. This
is difficult to see in panel ({\bf c}), but important because it allows trajectories approaching from above to
hit the collision sphere from below. The same conclusion holds for the invariant manifolds of  IV$'$.  In other words, the separatrices between colliding and non-colliding trajectories from above are no longer grazing trajectories,
but the stable manifolds of IV and IV$^\prime$.
As in panel ({\bf b}), there are closed trajectories  in the small regions delineated by the stable and unstable manifolds of IV~and~IV$^\prime$. These  trajectories are shown as blue dotted lines. 
 
Panel ({\textbf d}) corresponds to $Q=8.09$ and ${\rm Kn}=10^{-3}$. We see that the saddle IV and its symmetric counterpart IV$^\prime$  collided, giving rise to two new saddles, 
V and VI.  Now the collision sphere is entirely shielded from below by the unstable manifolds of VI.  The stable manifolds of V extend to $R_3 \to +\infty$ without touching the collision sphere. So again, some trajectories coming from $R_3=\infty$ can collide from below. Finally, we see that the fixed points II, II$^\prime$ merged with III which turned into  a stable node.

In summary, the phase portraits in Fig.~\ref{fig:phase1} are qualitatively different, and we expect that the qualitative changes to the fixed points and their manifolds affect how the collision rate depends on $Q$ and upon the Knudsen number Kn. To explain these dependencies, we must first locate and characterise the bifurcations of the collision dynamics.

\subsection{Bifurcations}
The dynamical system, Eq.~\eqref{eq:eom}, exhibits three different types of bifurcations when the parameters $Q$ and Kn are varied.
First, there are bifurcations of equilibria (labeled \textcircled{\footnotesize A} -  \textcircled{\footnotesize E} below), where the number, or types of equilibria change. Those bifurcations that occur at $R>2$ are smooth,  they must have the standard normal
forms of smooth dynamical systems \cite{Strogatz2000,Ott02}. 
But bifurcations at $R=2$ need not be smooth, because the radial derivatives 
of the mobilities diverge as $R\to 2$, as mentioned above.
Third,  grazing bifurcations (labeled \textcircled{\footnotesize $\alpha$} -  \textcircled{\footnotesize $\gamma$} below) are global bifurcations, where  an invariant manifold, such as a separatrix, touches the collision sphere. 

We begin by discussing the bifurcations of equilibria. Fig.~\ref{fig:phase1}({\bf e},{\bf f}) shows  how their equilibrium-locations $[R^\ast,\theta^\ast]$ change
as $Q$ is varied. 
The first bifurcation, \textcircled{\footnotesize A}, occurs at $Q_c=5.46$. It is a saddle-node bifurcation where the saddle IV and the node N are created, as well as their
symmetric counterparts IV$^\prime$ and N$^\prime$. For ${\rm Kn}=10^{-3}$, this occurs at $R^\ast-2 \approx 0.011$, quite close to the collision sphere.  Nevertheless, since the dynamical system is smooth for $R>2$, this is a standard smooth bifurcation. The location of the node is most clearly seen in the logarithmic scale of the inset in panel ({\bf e}).

The next bifurcation, \textcircled{\footnotesize B}, is at $Q_c=6.67$ where the nodes N and N$'$ collide with the saddles II and II$^\prime$ on the collision sphere, and  II and II$^\prime$ change from saddles to stable nodes. This bifurcation cannot occur in smooth dynamical systems, because 
in the absence of boundaries one can show that
the sum of  Poincar\'e indices of the fixed points participating in a bifurcation does not change  \cite{Strogatz2000}. 
We note that the Poincar\'e{}-index argument
is readily extended to systems with boundaries: enclose the fixed points II and N by a curve that starts and ends on the collision sphere,  infinitesimally close to II.  Since the saddle II changes into a node, the angle of rotation of the velocity field along this curve is conserved before and after bifurcation \textcircled{\footnotesize B}.

Numerical integration becomes difficult as N and N$'$ approach the collision sphere. We therefore sought an asymptotic approximation describing
how the location $R^\ast$ of the nodes approach the collision sphere as  $Q$ tends to the bifurcation value $Q_c^{\footnotesize \textcircled{\tiny B}}$.
We found
\begin{equation} \label{eq:N_asymptote}
 R^\ast \sim 2+{\rm exp}\big[-20.49/(Q_c^{\footnotesize \textcircled{\tiny B}}-Q)
 \big]\,.
\end{equation}
for $Q < Q_c^{\footnotesize \textcircled{\tiny B}}$.
Details of the derivation are given in Appendix~\ref{sec:appendix_b}, which also contains
a more accurate albeit more complicated expression [Eq.~(\ref{eq:B9})], plotted in Fig.~\ref{fig:phase1}({\bf e},{\bf f}) as red lines.
The non-analytic dependence on the parameter $Q$ in Eq.~\eqref{eq:N_asymptote} is a result of 
the divergence of the radial derivatives of the tangential mobility. In smooth systems, by contrast,   bifurcation locations
depend algebraically on $Q_c^{\footnotesize \textcircled{\tiny B}}-Q$ \cite{Strogatz2000}.
Eq.~(\ref{eq:N_asymptote}) implies that $R^\ast-2$ is almost
zero long before $Q$ has reached its bifurcation value $Q_c^{\footnotesize \textcircled{\tiny B}}$. 

Bifurcation \textcircled{\footnotesize C} happens at $Q=7.88$, where the saddles IV and IV$^\prime$ collide to form two new saddles, V and VI. Since this bifurcation occurs at $R^\ast> 2$, it
is a smooth bifurcation. At the bifurcation, two stable and two unstable manifolds of fixed points IV and IV$^\prime$ merge into a stable and an unstable one.
Both are oriented along $\hat{\bf e}_3$. The fixed point  therefore has six invariant manifolds at the bifurcation: three stable and three unstable ones. 
After the bifurcation, the saddle points split along $\hat{\bf e}_3$, connected by a new heteroclinic trajectory.

At $Q=8.04$, the fixed points II and II$'$ collide with III in a supercritical pitch-fork bifurcation \textcircled{\footnotesize D} (see Appendix~\ref{sec:appendix_b}), leaving
behind only III which turned from a saddle point to a stable node. This bifurcation occurs at $R=2$, but it is nevertheless smooth. The reason
is that the fixed points move along the collision sphere, and the corresponding one-dimensional dynamical system is smooth. 
The non-smooth behaviour of the tangential mobilities does not qualitatively alter the bifurcation.

The last bifurcation of an equilibrium, \textcircled{\footnotesize E}, occurs at $Q_c=10.41$ where  the saddle point V collides with the node III on the collision sphere,
which turns into a saddle point. This bifurcation cannot occur in smooth systems, where a saddle-node collision cannot result in a single saddle, 
as explained above.
Note that the results shown in panels~({\bf e},{\bf f})  are independent of the Knudsen number.
 The locations of  the fixed points I, II, II$^\prime$, and III depend only on tangential mobilities which do not depend upon Kn
in the model considered here. The other fixed points depend upon the radial mobilities only through the combination $L/(1-A)$, which is independent of the Knudsen number, 
as explained in Appendix~\ref{sec:appendix_b}.

In addition to these bifurcations of equilibria, there are three grazing bifurcations, labeled with Greek letters. Their locations are not shown in Fig.~\ref{fig:phase1}({\bf e,f}) because they
are global bifurcations not described by changes in equilibria.
For Kn=$10^{-3}$, the grazing bifurcation \textcircled{\footnotesize $\alpha$} occurs at $Q_c=5.67$, larger than
but quite close to $Q_c^{\footnotesize\textcircled{\tiny A}}$. At this bifurcation, the unstable manifolds of the saddles IV and IV$^\prime$ approaching the collision sphere 
end up intersecting it. The points of intersection are shown as circles in panel~({\bf b}).

The grazing bifurcation  \textcircled{\footnotesize $\beta$} occurs at $Q_c^{\footnotesize\textcircled{\tiny$\beta$}}=7.6$ for Kn=$10^{-3}$, when the  stable manifolds of IV and IV$'$ 
approaching from  $R_3=\infty$ graze the collision sphere, allowing paths from above to hit the collision sphere from below for $Q>Q_c^{\footnotesize\textcircled{\tiny$\beta$}}$.
A final grazing bifurcation, \textcircled{\footnotesize $\gamma$},  occurs at $Q_c^{\footnotesize\textcircled{\tiny $\gamma$}}= 8.1$ where  stable manifolds of the saddle V begin to graze the collision sphere, preventing trajectories approaching from above from colliding from below. 

The locations of the grazing bifurcations depend on the Knudsen number. As Kn$\;\to0$,  we find that $Q_c^{\footnotesize \textcircled{\tiny $\alpha$}}, Q_c^{\footnotesize \textcircled{\tiny $\beta$}} \to Q_c^{\footnotesize \textcircled{\tiny B}}$, while $Q_c^{\footnotesize \textcircled{\tiny $\gamma$}} \to Q_c^{\footnotesize \textcircled{\tiny E}}$. To see this, recall that  invariant manifolds begin to intersect the collision sphere at these grazing bifurcations.  This cannot happen for Kn$\,=0$ where  the radial mobilities 
vanish linearly in $\xi\equiv R-2$ as $\xi\to 0$.
In this limit, the  grazing bifurcations must therefore coincide with bifurcations \textcircled{\footnotesize B} and \textcircled{\footnotesize E} where a stable node appears or disappears at $R=2$.

\subsection{Collision rate}
What are the consequences of these bifurcations for the collision rate? 
 Fig.~\ref{fig:phase1}({\bf g}) shows the steady-state collision rate $\mathscr{R}_\infty=\lim_{t\to\infty} \mathscr{R}_t$  for $a_1/a_2=0.9$ as a function
of the non-dimensional settling speed $Q$, for three different values of the Knudsen number:
Kn $=10^{-1}, 10^{-2}$ and $10^{-3}$. Since the phase portraits in Figure~\ref{fig:phase1} suggest different behaviours for
paths approaching from below and from above, the collision rate was computed separately for the two cases.

When the separatrices delineating collisions from no collisions are grazing trajectories, as in Fig.~\ref{fig:phase1}({\bf a}), the collision rate decreases as Kn decreases, as seen in Fig.~\ref{fig:phase1}({\bf g}) for  $Q<Q_c^{\footnotesize \textcircled{\tiny $\alpha$}} $. The reason is that as the Kn number decreases, fewer trajectories approaching from large separations collide. However, Fig.~\ref{fig:phase1}({\bf g}) 
shows that the collision rate  becomes independent of Kn  for trajectories approaching from below, for small Kn, and for a certain
range of non-dimensional settling speeds $Q$ (between bifurcations \textcircled{\footnotesize $\alpha$} and \textcircled{\footnotesize C}).
This means that the collision rate does not vanish in this range for small Knudsen numbers. 

The bifurcation analysis summarised above explains why the collision rate remains non-zero.
The argument goes as follows.
Between bifurcations \textcircled{\footnotesize A} and \textcircled{\footnotesize $\alpha$}, there is a stable node at $R>2$. In this regime, collision outcomes for trajectories from above are  determined by a grazing trajectory, and so $\mathscr{R}$ depends on Kn. The fate of trajectories approaching from below is more subtle. 
Trajectories approaching from below between the stable manifolds of IV and IV$^\prime$ may either collide 
or asymptote to the stable node. Again, whether the trajectories collide or miss and approach the node, depends on the Knudsen number.
However, at bifurcation \textcircled{\footnotesize $\alpha$}, the invariant manifold connecting the stable node N to the saddle IV hits the collision sphere. As a consequence, all trajectories approaching from below between the stable manifolds of saddles IV and IV$^\prime$ must collide. Since these fixed points and their stable  manifolds 
are far from the collision sphere,  for the Knudsen numbers considered, their location does not depend on Kn. This implies, in turn, that the collision rate is independent of Kn
for trajectories approaching from below. This mechanism is analogous to that determining collisions of oppositely charged droplets settling in a quiescent fluid \cite{magnusson2021collisions}, where the separatrices are also formed by the stable manifolds of a saddle point of the collision dynamics. The former occurs when differential settling due to gravity and Coulomb attraction cancel. Thus the two bifurcations  \textcircled{\footnotesize A} and  \textcircled{\footnotesize $\alpha$}, have important consequences for the collision rate: \textcircled{\footnotesize A} gives rise to the saddle IV whose stable manifold determines the collision rate for trajectories approaching from below, for $Q>Q_c^{\footnotesize \textcircled{\tiny $\alpha$}}$.

The change in the mechanism determining the collision rate from a grazing trajectory for $Q<Q_c^{\footnotesize \textcircled{\tiny $\alpha$}}$ to an invariant manifold for $Q>Q_c^{\footnotesize \textcircled{\tiny $\alpha$}}$ leads to a sensitive dependence of the collision rate on $Q$ near this bifurcation, seen as sharp peak near $Q=5.67$ for the green curve in Fig.~\ref{fig:phase1}({\bf g}). The reason is that the collision rate jumps  at $Q_c^{\footnotesize \textcircled{\tiny $\alpha$}}$ in the limit Kn$\;\to0$.
Before the bifurcation, the collision rate tends to 0 as Kn$\;\to0$, but after the bifurcation the collision rate asymptotes to a non-zero, Kn-independent limiting value.

Finally consider the  small non-monotonic bump in the collision rate for trajectories approaching from the upper half plane for very small Knudsen number, Kn$=10^{-3}$.
\citet{Dhanasekaran2021} pointed out that the bump is 
 due to trajectories  that encircle the collision sphere and collide from below. The Knudsen number needs to be small enough for this to occur, so that grazing separatrices, such as those in Fig.~\ref{fig:phase1}({\bf c}) can exist.  At Kn$\;=10^{-3}$, this gives a very small contribution to the collision rate. 
 Nevertheless, this mechanism leads to a non-zero collision rate in the limit Kn$\;\to0$ when $Q_c^{\footnotesize \textcircled{\tiny B}}<Q<Q_c^{\footnotesize \textcircled{\tiny E}}$,  for trajectories approaching from above.

\subsection{Spatial clustering}
The dynamics close to a saddle point can lead to an accumulation of trajectories -- spatial clustering -- if the stability exponents of the saddle point   satisfy certain constraints. 
To find these conditions, we start from the steady-state continuity equation,  obtained from Eq.~\eqref{eq:ce} by setting $\tfrac{\partial P}{\partial t} =0$. Now consider what happens close to the saddle point VI in Fig.~\ref{fig:phase1}(\textbf{d}).  Its stable and unstable directions are $\hat{\bf e}_3$ and $\hat{\bf e}_1$, respectively. 
At the fixed point, $\ve V=0$. In its vicinity we linearise the velocity field,  $V_1 = \lambda_+ R_1$ and $ V_3 = \lambda_- \delta R_3$, where $\lambda_+$ and $\lambda_-$ are positive and negative stability exponents of the saddle, and $\delta R_3 =R_3-R_3^\ast$.  
The resulting steady-state continuity equation can then be solved using separation of variables. We find:
\begin{equation} \label{eq:saddle_divergence}
 P({R_1,\delta R_3}) \sim   | \delta R_3|^{-\tfrac{\lambda_+}{\lambda_-}-1}  \; \text{as} \; {R_1\;\mbox{and}}\; \delta R_3 \to 0.
\end{equation}
We conclude that $P$ diverges algebraically as one approaches the unstable manifold in the vicinity of the fixed point, provided that
$0>\lambda_+/\lambda_- > -1$.

\begin{figure}[t]
\begin{overpic}[scale=0.2]{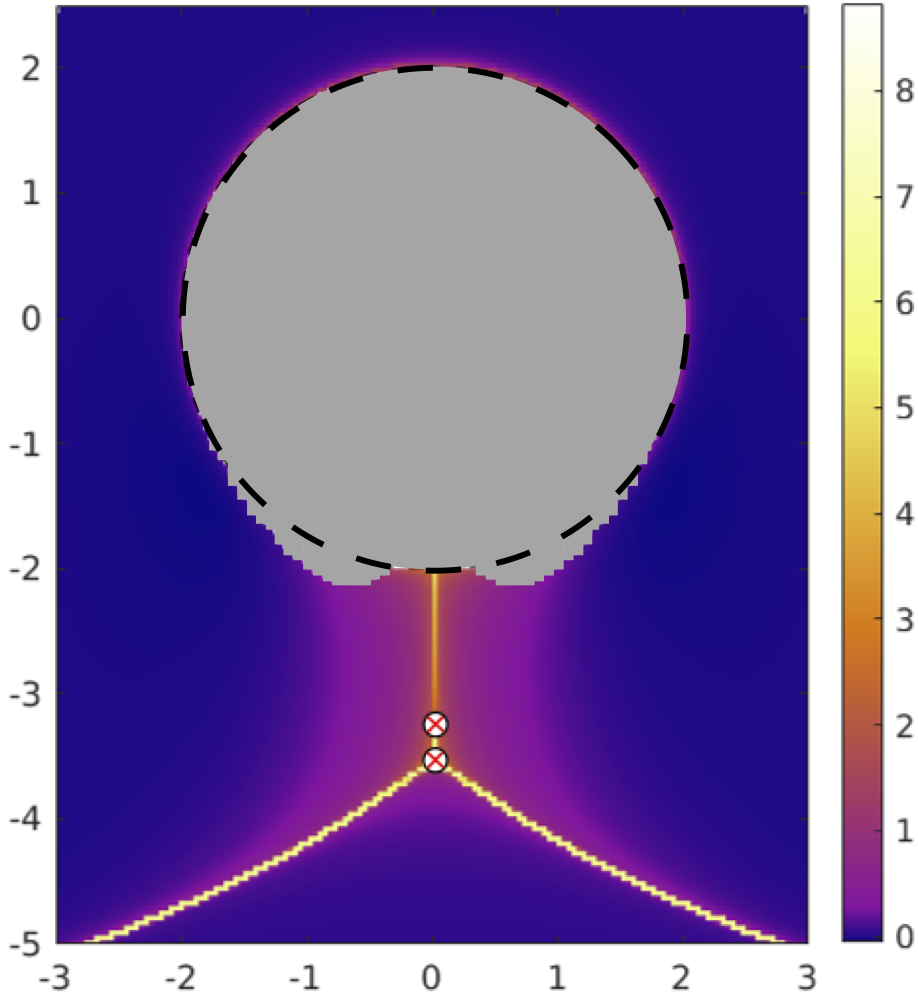}
\put(-3.5,73.5){$R_3$}
\put(59,-3){$R_1$}
\put(93,78.5){\rotatebox{90}{$\log_{10}P(\ve R)$}}
\end{overpic}
 \caption{\label{fig:density} Spatial clustering on invariant manifolds.
 Shown is the probability $P(\ve R)$ for $Q=8$, Kn$\,=10^{-3}$ in the $R_1$-$R_3$ plane, colour-coded on a logarithmic scale. The probability
 diverges along manifolds of fixed point VI [see Fig.~\ref{fig:phase1}({\bf d})]. The collision sphere is shown as a dashed circle. 
 The encircled red crosses shows the location of the saddle points V and VI. The gray regions cannot be reached from $R=\infty$.}
\end{figure}

Figure~\ref{fig:density} shows results for $P(\ve R)$ for  $Q=8$, $a_1/a_2=0.9$ and Kn=$10^{-3}$ obtained by numerically 
solving Eq.~\eqref{eq:ce} using the method of characteristics.
The saddle points V and VI are shown as encircled red crosses, just
 as in   Fig.~\ref{fig:phase1}. The phase portrait for $Q=8$  is similar to  panel ({\textbf d}) in that figure.
 For the saddle point VI, ${\lambda_+}/{\lambda_-} = -0.439$, and we observe 
regions of very large probability $P$ that coincide precisely with its unstable manifolds.
 For the saddle point V, by contrast ${\lambda_+}/{\lambda_-} = -1.23$. Since this ratio is smaller than $-1$,  no clustering near its unstable manifold is expected. The large probabilities observed along the line $R_1 = 0$ are due to a different mechanism, explained next.
  The probability $P$ diverges also very close to the collision sphere. This divergence, hard to see in Fig.~\ref{fig:density} because it occurs so close to the collision sphere
  and because it is much weaker, was discussed
in Refs.~\cite{Dhanasekaran2021,Dhanasekaran2021b}. 
Since the  radial relative velocity decays to zero as $V_R \sim (\log \log \tfrac{\text Kn}{{R-2}})^{-1}$ as $R\to2$, the steady-state continuity equation
implies that $P$  diverges as $\sim 1/V_R$ in this limit.

Near the saddle  VI, the divergence is stronger. Consider
the pair-correlation function $g(R)$. It is defined as the probability to find the two particles at separation $R$  \cite{Gus16},
 $g(R) = \langle P(\ve R)\rangle_R$. Here the average
 is over the polar angle $\theta$ at a fixed separation $R$.   Using Eq.~\eqref{eq:saddle_divergence}, we find that $g(R)$ exhibits an algebraic divergence as $R\to R^\ast$.

\section{Discussion}

We concluded above that the collision rate for trajectories approaching from below becomes independent of Kn for small Kn
between bifurcations \textcircled{\footnotesize $\alpha$} and \textcircled{{\footnotesize C}}. Since \textcircled{\footnotesize $\alpha$} is a grazing bifurcation, its location $Q_c^{\footnotesize \textcircled{\tiny $\alpha$}}$  must depend on Kn. 
Eq.~\eqref{eq:N_asymptote} shows that $R^\ast- 2 \sim 10^{-8}$  when $Q \approx 6$,
still very far from $Q_c^{\footnotesize \textcircled{\tiny B}}=6.67$.   In order to understand how \textcircled{\footnotesize $\alpha$} depends on Kn, consider how the invariant manifold connecting IV and N approaches the node. Its eigenvalues have non-zero imaginary parts, this means that the manifold
 passes  between the node and the collision sphere before  it reaches the node. 
 As a first rough estimate let us assume that this happens when $ R^\ast -2$ is of the order of Kn. Using this estimate and Eq.~\eqref{eq:N_asymptote}, we obtain $Q_c^{\footnotesize\textcircled{\tiny$\alpha$}}\sim Q_c^{\footnotesize\textcircled{\tiny B}}+ {20.49}/\log{\text{Kn}}$. This indicates that the Kn-derivative of $Q_c^{\footnotesize\textcircled{\tiny$\alpha$}}$  may diverge as Kn$\;\to0$, implying that $Q_c^{\footnotesize\textcircled{\tiny$\alpha$}}$ depends sensitively on the Knudsen number.

It is commonly stated that droplets cannot collide in the hydrodynamic approximation because it would take infinite time for them to touch \cite{Pru10,Pum16, Hocking1973, Davis1984}, 
 implying that the collision rate vanishes as ${\rm Kn}\to 0$. 
 However, we concluded here that  this is correct only when  the trajectory delineating collisions from no-collision is a grazing trajectory, as in Fig.~\ref{fig:phase1}(\textbf{a}). In this case, the flux of colliding trajectories approaching from afar tends to zero because the grazing trajectories touch at $R_3=\infty$ in the limit  Kn$\to0$, so that the steady-state collision rate from above  must vanish.  However, when collisions are determined by the invariant manifolds of saddle points, as for the trajectories approaching from below in Fig.~\ref{fig:phase1}(\textbf{b}), the flux of approaching trajectories does not depend on Kn if the distance between the stable manifolds  and the collision sphere is larger than Kn.  
  The time these trajectories take to collide 
depends weakly on the Knudsen number, $\propto \log(1/{\rm Kn})$. 
So the time it takes to collide grows as ${\rm Kn}\to 0$, but the steady-state collision rate remains
finite in this limit, because the number density of droplets builds up as their dynamics slows down.

Our bifurcation analysis explains closed trajectories in relative dynamics, first reported by \citet{Batchelor1972}, and later by \citet{Zeichner1977} and \citet{Dhanasekaran2021b}. These closed trajectories arise because some trajectories starting close to the stable node N [or close to the stable nodes II,II$'$ or III, depending on the value of $Q$] start and end on the collision sphere. At larger values of Kn, there are more such closed trajectories, because some paths that grazed the collision sphere at small Kn collide instead.  \citet{Zeichner1977} connected an observed decrease in collision rate to the appearance of such closed trajectories. Our results show, however, that is not the correct explanation, at least in the model considered here. The
 collision rate decreases between $ Q_c^{\footnotesize\textcircled{\tiny $\alpha$}}$ and $ Q_c^{\footnotesize\textcircled{\tiny C}}$ because the collision
 rate  for trajectories approaching from below decreases.  
 Since the closed trajectories do not contribute to collision rate for trajectories approaching from below,  they cannot explain the decrease of collision rate in this model.
 The correct explanation, outlined in the previous Section, is that the collision rate is determined by the stable invariant manifolds of the fixed points IV and IV$'$.

 Only the bifurcations  \textcircled{\footnotesize {A}}, \textcircled{\footnotesize {C}}, and \textcircled{\footnotesize {D}} are well explained by  bifurcation
 theory for local equilibria in smooth dynamical systems, because they occur at $R^\ast >2$ where the system is smooth. Bifurcations \textcircled{\footnotesize{B}} and \textcircled{\footnotesize {E}} occur at $R=2$ where the system is not smooth: the radial derivative of the tangential mobility  diverges. One  consequence is the non-analytic behaviour described by Eq.~\eqref{eq:N_asymptote}, very different from the algebraic scaling observed for smooth systems \cite{Strogatz2000}, and also 
 different from the behaviours found in piecewise smooth systems \cite{diBernardo2008, Brogliato2016}.

In our case, the collision dynamics stops at the boundary at $R=2$ because we imposed that  the droplets coalesce.  Recent experiments \cite{Bragg22} considered the collision dynamics of glass spheres in turbulence. Glass spheres may roll along the collision sphere at $R=2$, leading to new bifurcations \cite{Jeffrey2011}.
If the particles collided elastically instead, they could bounce. This might lead to interesting bifurcations similar to those seen in impacting systems \cite{Foale1994}. We note also that \citet{Bragg22}  observed pairs of spheres that stuck together. These could be
the result of inelastic collisions, corresponding to a third qualitatively different collision outcome. All of the three cases mentioned above may 
result in new bifurcations, and might even lead
 to chaotic dynamics~\cite{diBernardo2008}. 

In our analysis we considered only one value of the radius ratio, $a_1/a_2= 0.9$. Because the mobility functions are not sensitive to changes 
of this ratio when it is close to unity \cite{Jeffrey1984}, the bifurcation analysis should remain similar as the ratio becomes larger. 
The limit  $a_1/a_2\to 0$, by contrast, is expected to be quite different. 
 In this case, Kn-corrections
to the tangential mobilities qualitatively change the dynamics \cite{li2021non}.
How the phase-space portraits and the bifurcation diagrams change in this limit are open questions.

We assumed that  the strain is  aligned with gravity. \citet{Dhanasekaran2021} studied numerically how the collision rate changes when the angle between strain and gravity changes. Their results indicate that the peak in the collision rate at $Q\approx6 $ is instead replaced by a minimum as the angle between strain and gravity changes. How are phase-space portraits affected, and what are the corresponding changes to the bifurcation diagram? 
When strain and gravity is aligned, the dynamics is rotationally invariant. As a consequence, there is a continuous circle of fixed points \cite{Strogatz2000},
of which we considered only one pair, IV and IV$^\prime$. For non-zero angles, this circle breaks up. Numerical simulations show that the qualitative conclusions regarding how the invariant manifolds determine collision outcomes
remain similar for small but non-zero angles. A question for the future is to locate and characterise the new bifurcations that  must occur as the angle is varied, and to analyse
their effect upon the collision rate. 

The  spatial clustering we described is very anisotropic because it occurs for large values of the non-dimensional differential settling speed $Q$ where gravity dominates. Another important point is that the probability diverges not
only close to the particle, but also at separations of the order of a few droplet radii.  As explained above, this is a consequence of the existence of saddle points in the relative dynamics.
An open question is how misalignment between strain and gravity affects the spatial clustering. 
 It is likely that an anisotropy persists if one averages over  randomly oriented strains, because gravity breaks up-down symmetry. 
 But the details remain
 to be worked out. 
 
 The results reported here concern very small, inertialess droplets. How do the phase portraits change when particle inertia is included? First, the full phase-space dynamics including  velocities must be considered, as described in Ref.~\cite{magnusson2021collisions}. Second,  particle inertia allows the particles to overshoot the saddle point IV in Figure~\ref{fig:density}.
 This could reduce the spatial clustering.
 Third, while in the inertialess case the relative velocity at contact  must vanish, inertia allows collisions at non-zero relative velocities \cite{Gus16}, causing the collision rate to increase. 
 An example is given by \citet{Bec2020} who describe how particle inertia increases the collision rate of small particles colliding with a large one. Furthermore, for solid spheres, collisions at non-zero relative velocities could give rise to chaotic dynamics, similar to  transitions to chaos observed in impacting systems \cite{diBernardo2008}.

Finally, our analysis considered steady flow.
In time-dependent flow, such as turbulence, the bifurcation analysis becomes more difficult since one has to consider
a driven dynamical system.
 In particular it remains to be seen to which extent
the  spatial clustering is weakened when the flow becomes time dependent. We note that recent experimental studies \cite{Yav18,Bragg22} of colliding particle
in turbulence show strong spatial clustering at small separations. Whether or not the clustering mechanism described above can offer an explanation of these observations, depends on
whether it survives averaging over different steady linear flows, and to which extent it is weakened by stochastic driving in a time-dependent flow.

\section{Conclusions}
We investigated bifurcations in the collision dynamics of small droplets settling in a straining flow. We considered the case
where the compressive direction of strain is aligned with gravity. We showed that the collision dynamics is not smooth because the
radial derivatives of the mobility functions diverge on the collision sphere. As a consequence, we observed types of bifurcations  that
cannot occur in smooth dynamical systems where index theorems constrain the possible normal bifurcation forms. We found three different kinds of bifurcations: standard smooth bifurcations of equilibria, but also non-smooth bifurcations of equilibria, as well as grazing bifurcations.

The grazing bifurcations determine  the global phase portraits of the collision dynamics and explain
that the collision rate approaches a non-zero constant for small Kn, for a range of non-dimensional settling velocities.
This is surprising, because it is commonly stated that droplets cannot collide in the hydrodynamic approximation \cite{Hocking1973, Davis1984,Pru10, Pum16}, implying that the
steady-state collision-rate vanishes. Our results show that
this argument is correct only when the trajectories separating colliding from non-colliding paths graze the collision sphere. This is the case for droplets with different radii
settling in a quiescent fluid.
But when differential settling competes with a straining flow, we concluded here  that these separatrices are, in a certain parameter range, formed by the stable manifolds of saddle points that are far from the collision sphere. In this case the steady-state collision 
rate becomes  independent of the Knudsen number for small Kn.

Our analysis shows that saddle points may give rise to  spatial clustering, causing trajectories to accumulate close to the saddle point in question,
leading to a flux along the unstable manifold of the saddle. We demonstrated that this can result in an algebraic divergence
of the pair correlation function $g(R)$ at separations $R$ of the order of several droplet radii.
 Since gravity breaks the up-down symmetry of the problem, the observed spatial clustering is highly anisotropic.  When the compressive  axis of the straining flow is aligned with gravity, there is a high probability of observing the larger sphere  below the smaller one. This  prediction may be tested experimentally by looking at snapshots of droplets in gravity-dominated straining flow.
 
Our results show that understanding smooth and non-smooth bifurcations of the collision dynamics yields important insight into the physics of the collision process. 
In the future we therefore intend to perform this analysis for more general cases. The first step is to consider droplet radius ratios other than $a_1/a_2=0.9$.  We expect that the results remain similar
when the ratio is larger than this value. But when  $a_1/a_2\to 0$, the model considered here may fail because it neglects Kn-corrections to the tangential mobilities
which may become important in this limit. 

Second, we intend to investigate the bifurcations when strain is not aligned with gravity. 

Third, in order to describe droplet collisions in turbulence, we need to consider what happens when the flow is time dependent. In order to understand how the collision rate depends on Kn, the timescales at which stable manifolds of saddle points move have to be considered. Moreover, it is unclear to which extent spatial clustering survives in this case. If it does, it could be a mechanism for the extreme spatial clustering observed in recent experiments  \cite{Yav18,Bragg22}, which remains unexplained. 

Fourth, particle inertia matters for larger droplets, because it allows them to detach from the flow \cite{Gus16}. The bifurcation analysis in this case is more challenging because it must be performed in phase space including separations and relative velocities.  

Finally, 
in an earlier study  \cite{magnusson2021collisions}, we analysed the effect of electrical charges on the collision dynamics. Ref.~\cite{magnusson2021collisions}
did not account for the breakdown of the hydrodynamic  approximation close to the collision sphere,
therefore we considered charges large enough that the relevant equilibria were far from the collision sphere. In this limit we found the phase-portraits to be qualitatively different
from those for neutral droplets. But we do not know at which amount of charging this qualitative change happens. It might occur already for small
charges of the order of typical charges observed in warm rain clouds \cite{Tak73}. This could have important implications for droplet collisions in rain clouds.

{\em Acknowledgements}.
 The research of AD and BM was supported by grants from Vetenskapsr\aa{}det (grant nos.~2017-3865 and 2021-4452 and from the Knut and Alice Wallenberg Foundation (grant no.~2014.0048). The research of GB was supported by the National Science Foundation under grant no.~CBET-1605195. BM acknowledges a  Mary Shepard B. Upson Visiting Professorship with the Sibley School of Mechanical and Aerospace Engineering at Cornell.

\appendix
\section{Fixed-point locations of  I-III,  and analysis of bifurcation \textcircled{\footnotesize{D}}} \label{sec:appendix_a}

Eq.~\eqref{eq:eom} in the main text can be expressed in radial and angular components of relative velocity, $V_R \equiv \dot R$ and $V_\theta \equiv R \, \dot \theta$ as,
\begin{subequations}
\label{eq:radial}
\begin{align}
 V_R &\equiv - L Q \cos \theta + (A-1) \frac{R}{4} (1+3 \cos 2 \theta)\,, \\
 V_\theta &\equiv  M Q \sin \theta + (1-B) \frac{3R}{4} \sin 2 \theta \,.
\end{align}
\end{subequations}
Here $L$ and $M$ are the radial and tangential mobility functions for spheres due to sedimentation, while $A$ and $B$ are the radial and tangential mobility functions for spheres in  shear flow \cite{Batchelor1972, Batchelor1982}.

Fixed points may be found by solving the equations $V_R =0 = V_\theta
$. The fixed points I, II, II$^\prime$, and III lie on the collision sphere where $V_R=0$. 
In order to find these fixed points, we must solve $V_\theta = 0$ at $R=2$,
 \begin{equation} \label{eq:bifurcation_D}
  [M_0 Q - 3 B_0 \cos \theta]\sin \theta =0\,.
 \end{equation}
 Here $B_0$ and $M_0$ are the values of the tangential mobility functions upon contact.
Eq.~(\ref{eq:bifurcation_D}) has solutions $\theta^\ast = 0$  and $\pi$, corresponding to fixed points I and III, respectively.  In addition, if $Q\leq 3\,B_0/M_0$ the equation has solutions $\theta^\ast  = \pm\arccos \tfrac{M_0 Q}{3 B_0}$. These correspond to fixed points II and II$^\prime$. As $Q$ increases from zero, the fixed points II and II$^\prime$ move along the collision sphere until $Q_c^{\footnotesize \textcircled{\tiny D}}= 3\,B_0/M_0$ when they collide with III. This is bifurcation \textcircled{\footnotesize{D}}, a
supercritical pitchfork bifurcation  \cite{Strogatz2000}. For $Q>Q_c^{\footnotesize \textcircled{\tiny D}}$, the equation $\cos \theta = \tfrac{M_0 Q}{3 B_0}$ admits no real solutions.

\section{Fixed-point location of N, and analysis of bifurcation \textcircled{\footnotesize{B}}} \label{sec:appendix_b}

In order to compute the   location $[R^\ast,\theta^\ast]$ of the node N, we solve $V_R =0$ for the fixed-point angle,
\begin{subequations}
 \label{eq:appendix_theta0}
\begin{align}
\label{eq:thetaast}
 \theta^\ast &= \arccos \Big( \frac{-4k -\sqrt{k^2 +48}}{12}\Big)
 \end{align}
 with
 \begin{align}
 \label{eq:k_def}
 k = \tfrac{\, L\, Q}{(1-A) \, R^\ast} \,.
\end{align}
\end{subequations}
Inserting this expression for $\theta^\ast$ into  $V_\theta =0$, we obtain
\begin{align} \label{eq:fp_solve}
 3(1-B) \frac{R^\ast}{4} \sin 2\theta^\ast +M Q \sin \theta^\ast = 0\,.
\end{align}
This is an implicit equation for $R^\ast$ because the mobility functions as well as $\theta^\ast$ [Eq.~\eqref{eq:appendix_theta0}] are  functions of  $R^\ast$. In general, this equation must be solved numerically. But asymptotic solutions can be obtained when $R^\ast$ is close to the collision sphere, 
using  known asymptotic expansions for the mobilities for small $\xi \equiv R-2$.

The radial mobility functions $A$ and $L$ depend upon the Knudsen number only through the functions $\Lambda_{11}^{nc}$ and $\Lambda_{21}^{nc}$ defined in Equations (4.2)--(4.6) in Ref.~\cite{Dhanasekaran2021}. From their Equations (4.2)--(4.3), it follows that the ratio $L/(1-A)$ is independent of both $\Lambda_{11}^{nc}$ and $\Lambda_{21}^{nc}$. Thus, the ratio $L/(1-A)$ does not depend upon the Knudsen number. For small $\xi$, this ratio approaches the constant \cite{Batchelor1972, Batchelor1982, Jeffrey1984}:
\begin{align}
\label{eq:ratio0}
 \frac{L(\xi)}{1-A(\xi)} \sim \frac{1}{C}  \,.
\end{align}
Eq.~(\ref{eq:ratio0}) allows us to  approximate  $k$ in Eq.~\eqref{eq:k_def} as
\begin{align} \label{eq:asymptotic_beta}
 k \sim \frac{ Q}{{C}\, (2+\xi)} \quad\mbox{for}\quad \xi \ll 1\,.
\end{align}
The asymptotic expansions for the tangential mobilities in the limit $\xi \to 0$ read \cite{Jeffrey1984}
\begin{align} \label{eq:appendix_tangential}
B,M \sim \frac{c_{B,M}^{(1)} \log^2 \xi-c_{B,M}^{(2)} \,\log \xi+c_{B,M}^{(3)} }{\log^2 \xi-e^{(1)} \,\log \xi+e^{(2)}}\,.
\end{align}
The coefficients $c_{B,M}^{(i)}$ and $e^{(i)}$ depend only on the radius ratio $a_1/a_2$. While the coefficients $c_{B,M}^{(i)}$ in the numerator are different for the
 two mobility functions $B$ and $M$, the coefficients $e^{(1)},e^{(2)}$  in the denominator are the same for these functions.

We now insert Eqs.~\eqref{eq:appendix_tangential},    \eqref{eq:asymptotic_beta},  as well as \eqref{eq:thetaast}, into 
Eq.~\eqref{eq:fp_solve}, obtaining an equation for the fixed-point location 
valid for small $\xi$ and $\eta \equiv -1/\log \xi$. In order to find its asymptotic solutions in the limit $\xi \to 0$, note that order-$\xi$ terms are sub-leading to order-$\eta$ terms. Ignoring the former, one obtains a quadratic equation in $\eta$. The solution yields $\eta$ as a function of $\delta Q= Q_c^{\footnotesize \textcircled{\tiny B}}-Q$,
\begin{subequations}
\begin{align}
 \eta&=f(\delta Q),\\
 Q_c^{\footnotesize \textcircled{\tiny B}}&= B_0\sqrt{\frac{3{C} }{M_0({C}M_0-B_0)}} .
\end{align}
\end{subequations}
 The function $f$ has a zero at $Q=Q_c^{\footnotesize \textcircled{\tiny B}}$, corresponding to the value of $Q$ at which the node N touches the collision sphere. This corresponds to bifurcation \textcircled{\footnotesize B}. For $a_1/a_2=0.9$, we obtain $Q_c^{\footnotesize \textcircled{\tiny B}} = 6.6759$, and the function $f$ has the series expansion 
 \begin{align}
 f(\delta Q) = 0.0488\, \delta Q -0.0255 \, \delta Q^2 
 +0.0147\, \delta Q^2 +\dots
\end{align}
in $\delta Q= Q_c^{\footnotesize \textcircled{\tiny B}}-Q$.
Changing variables from $\eta$ back to $\xi$ we arrive at
\begin{align}
\label{eq:B9}
 \xi = \exp  [-1/f(\delta Q) ]\,.
\end{align}
Inserting $f(\delta Q)\approx  0.0488\, \delta Q$ yields Eq.~\eqref{eq:N_asymptote} in the main text.

\end{document}